\documentstyle[preprint,aps]{revtex}
\newcommand{\be}{\begin{eqnarray}}
\newcommand{\ee}{\end{eqnarray}}
\begin{document}
\title{Vortex Velocity Probability Distributions in 
Phase Ordering Kinetics}
\author{Gene F. Mazenko}
\address{The James Franck Institute and Department of Physics}
\address{The University of Chicago, Chicago, Illinois 60637}
\date{\today}
\maketitle

\begin{abstract}

The calculation of the point vortex
velocity probability distribution function (vvpdf)
is extended to a larger class of systems beyond
the nonconserved TDGL model treated earlier.  
The range is extended
to include certain anisotropic models and the 
conserved order parameter case.   The vvpdf still 
satisfies scaling with large velocity tails
as for the nonconserved isotropic case.
It is shown 
that the average vortex speed can be self-consistently
expressed in
terms of correlation functions associated with a
Gaussian auxiliary field.  In the conserved order 
parameter case the average vortex speed decays as
$t^{-1}$ compared to the $t^{-1/2}$ decay for the
nonconserved case.

\end{abstract}

\pacs{PACS numbers: 05.70.Ln, 64.60.Cn, 64.60.My, 64.75.+g}

\section{Introduction}
In recent work\cite{QM03} it was shown that the
theoretically predicted\cite{mazVV}  velocity probability
distribution for point vortices in the case of a phase ordering
system agrees very well with direct numerical simulations.
In particular the predicted high velocity algebraic tail
is found to be robust and the predicted exponent confirmed.
In the original paper\cite{mazVV} describing the theory there were
assumptions concerning the Gaussian nature of an underlying
auxiliary field.  
Here we clarify this result by showing that the assumption of
an underlying Gaussian field is consistent and does not
imply that the underlying order parameter field
is Gaussian.  We only require that the order parameter
field and the auxiliary field share the same zeros and
symmetry.
The theory is also extended here to conserved and anisotropic
systems of coarsening point defects.

In ref.\onlinecite{mazVV} it was shown for a nonconserved
time-dependent Ginzburg-Landau (TDGL) model
that for annihilating
point defects $n=d$, where $n$ is the number of components of
the order parameter and $d$ the spatial dimensionality of the
system, that the vortex velocity probability distribution
function (vvpdf), the probability that a given vortex has
the velocity ${\bf V}$ at 
time t  after a quench, is given by
\be
P({\bf V},t)=\frac{\Gamma (\frac{n}{2}+1)}{(\pi \bar{v}^{2}(t))^{n/2}}
\frac{1}{\Biggl( 1+{\bf V}^{2}/\bar{v}^{2}(t)\Biggr)^{(n+2)/2}}
\label{eq:1}
\ee
where the parameter $\bar{v}(t)$ is clearly related to the 
average vortex speed and varies as
$t^{-1/2}$ for long times for the nonconserved TDGL model.
Both the form of $P({\bf V},t)$ and the time dependence of
$\bar{v}(t)$ have been confirmed in 
ref.\onlinecite{QM03} for the case $n=d=2$.
It is worth pointing out that the order parameter growth law,
$L(t)$,
for the system studied in Ref.\onlinecite{QM03}
has a log correction\cite{YPKH,PFGPYC},
$L^{2}\approx t/ln~t$.  This log correction for
$L(t)$ is seen in the simulations Ref.\onlinecite{QM03}.
There are no log corrections found for
$\bar{v}(t)$.  Thus nonlinearities which influence
$L$ are not seen in $\bar{v}(t)$.
We discuss in more detail here the derivation of the 
result given by Eq.(\ref{eq:1})  
and it's extension to systems with a conserved order parameter
and simple spatial anisotropy.  
In the end we find that the vvpdf still satisfies a
form similar to Eq.(\ref{eq:1}) with the same large
velocity tail, but the average vortex speed falls
off as $t^{-1}$ in the conserved case compared to the
$t^{-1/2}$ behavior found for the nonconserved case.

The set of problems of interest here are driven by Langevin
equations of the form
\be
\frac{\partial \psi_{\alpha}}{\partial t}=
K_{\alpha}(\psi )
\ee
where, we assume,
\be
\lim_{\psi\rightarrow 0} K_{\alpha}(\psi )
=-\hat{O}\psi_{\alpha}
\label{eq:5}
~~~,
\ee
and the right-hand side is linear in $\psi$.
The key idea is that we are interested in that part of
the equation of motion which corresponds to the motion of
vortex cores which are characterized by zeros of the 
order parameter.  An important example is the TDGL 
model which is of the form
\be
\frac{\partial \psi_{\alpha}}{\partial t}=
-\hat{\Gamma} \left[V_{\alpha}^{\prime}(\psi )
-c\nabla^{2}\psi_{\alpha}\right]
\label{eq:6}
\ee
where $c > 0$,
$\hat{\Gamma}$ is a constant for a nonconserved
order parameter (NCOP) and
$\hat{\Gamma}=-D\nabla^{2}$
for a conserved order parameter (COP).  Comparing
Eq.(\ref{eq:6}) and Eq.(\ref{eq:5}) we have
\be
\hat{O}_{NCOP}(1)=-\Gamma c \nabla^{2}_{1}
\ee
and
\be
\hat{O}_{COP}(1)=D \nabla^{2}_{1} [-r+c\nabla^{2}_{1}]
\ee
for the COP case.  Here $r=V''(\psi )|_{\psi =0}$
and $r < 0$ if the system is unstable.
Through a proper choice of length 
and time scales we can choose
\be
\hat{O}_{NCOP}(1)=-\nabla^{2}_{1} 
\ee
\be
\hat{O}_{COP}(1)=\nabla^{2}_{1} [1+\nabla^{2}_{1}]
~~~.
\ee
An anisotropic model can be assumed
to be of the form
\be
\hat{O}_{ANI}=-\sum_{\mu_{1}}c_{\mu_{1}}\nabla_{\mu_{1}}^{2}
+\sum_{\mu_{1},\mu_{2}}b_{\mu_{1},\mu_{2}}\nabla_{\mu_{1}}^{2}\nabla_{\mu_{2}}^{2}
\ee
where $c_{\mu_{1}}$ and $b_{\mu_{1},\mu_{2}}$ are constants.
This reduces to the COP case if $c_{\mu_{1}}=-1$ and
$b_{\mu_{1},\mu_{2}}=1$.

There is the underlying assumption that the nonlinear potential
contribution in the equation of motion
must be such that system orders via
annihilating point defects.  

\section{Defect Densities and Continuity Equation}

We assume that the
instantaneous positions of these defects are
determined by the zeros of the order parameter field.
Furthermore, it was pointed out in ref.\onlinecite{mazVV},
that the vortex charge density for this system can be written as
$\rho =\delta (\vec{\psi}){\cal D}$
where ${\cal D}$ is the Jacobian (determinant) for the change of
variables from the set of vortex positions $r_{i}(t)$
(where $\vec{\psi}$ vanishes)  to the field
$\vec{\psi}$:
\be
{\cal D}=\frac{1}{n!}\epsilon_{\mu_{1},\mu_{2},...,\mu_{n}}
\epsilon_{\nu_{1},\nu_{2},...,\nu_{n}}
\nabla_{\mu_{1}}\psi_{\nu_{1}}
\nabla_{\mu_{2}}\psi_{\nu_{2}}....
\nabla_{\mu_{n}}\psi_{\nu_{n}}
\ee
where $\epsilon_{\mu_{1},\mu_{2},...,\mu_{n}}$ is the
$n$-dimensional fully anti-symmetric tensor and
summation over repeated indices here and below is implied.
Furthermore, since topological charge is conserved,
it was shown in ref.\onlinecite{mazVV} that $\rho$
satisfies a continuity equation:
\be
\frac{\partial\rho}{\partial t}
=-\vec{\nabla}\cdot\left(\rho {\bf v}\right)
\label{eq:14}
\ee
where the vortex velocity is given by
\be
{\cal D}v_{\beta}
=-\frac{1}{(n-1)!}\epsilon_{\beta,\mu_{2},...,\mu_{n}}
\epsilon_{\nu_{1},\nu_{2},...,\nu_{n}}
(\hat{O}\psi_{\nu_{1}})
\nabla_{\mu_{2}}\psi_{\nu_{2}}....
\nabla_{\mu_{n}}\psi_{\nu_{n}}~~~.
\label{eq:10}
\ee
It is assumed that the velocity field is used inside expressions
multiplied by the vortex
locating  $\delta$-function so we can use Eq.(\ref{eq:5})
in Eq.(\ref{eq:10}).
These results are rather general. 

Notice that $\rho$ and ${\bf v}$ have certain important
invariance properties.  If we can write
\be
\psi_{\nu}(1)=\left(\alpha +\beta m^{2}(1)+\ldots
\right) m_{\nu}(1)
\label{eq:7a}
\ee
for small $m_{\nu}$ and 
where $\alpha$ and $\beta$ are constants,  it is easy to
see that
$\rho(\vec{\psi})=\rho(\vec{m})$.
If we further assume, as $\vec{m}$ and $\vec{\psi}$
go to zero,
\be
\hat{O}\psi_{\nu}(1)=\alpha\hat{O}m_{\nu}(1)
~~~,
\label{eq:7b}
\ee
then
\be
v_{\mu}(\vec{\psi}) =v_{\mu}(\vec{m})
~~~.
\label{eq:18}
\ee
If $\hat{O}$ corresponds to an operator
with two gradients, as in the NCOP TDGL
model, then, assuming Eq.(\ref{eq:7a}) to be
valid, Eq.(\ref{eq:7b}) follows.  In the case
where $\hat{O}$ has higher-order derivatives and we assume
Eq.(\ref{eq:7b}) holds, then Eq.(\ref{eq:7a}) must be
modified.

Thus the correlation function we compute in
the next section, $G(12)$, is for that set
of fields ${\bf m}$, related to to $\psi$ by
Eq.(\ref{eq:7a}), for small values of 
$m_{\nu}$ and  $\psi_{\nu}$,  which is described by a 
Gaussian distribution.  Thus we assume there is a field
$m_{\nu}$ which is Gaussian while the statistics
of $\psi_{\nu}$ are largely undetermined. 

\section{Correlations for the Defect Sector}

We show in Sect. IV that in determining
the average vortex speed we 
need certain correlation
functions for the
auxiliary field, ${\bf m}$, introduced in the last
section.  We show here that we can use the
defect continuity equation, Eq.(\ref{eq:14}), to show that there
is a self-consistent solution for an
${\bf m}$ field that is Gaussian. 
Furthermore we determine the
correlation functions needed to evaluate the
average vortex speed explicitly.
The method we develop here is a generalization
of the approach due to Mazenko and Wickham\cite{MW98}.

The idea is
to look at the equation generated by  multiplying
the continuity equation  by a source function and
then averaging over ${\bf m}$.  We have
\be
\langle \left[ {\partial\rho (1)}{\partial t_{1}}
+\vec{\nabla}^{(1)}\cdot
\left(\rho(1) {\bf v}(1)\right)
\right]S(H)\rangle =0
\label{eq:8}
\ee
where
$S(H)=exp[\int~d\bar{1}
{\bf H}(\bar{1})\cdot{\bf m}(\bar{1})]$.
The question is whether this equation can be satisfied
for an underlying Gaussian probability distribution
for arbitrary external field ${\bf H}(1)$?

To answer this question we evaluate first the quantity
\be
\langle \rho (1)S(H)\rangle
=\langle S(H)\delta (1){\cal D}(1)\rangle
\label{eq:15a}
\ee
where we introduce the simplifying notation
$\delta (1)=\delta ({\bf m}(1))$
and now ${\cal D}$ is a function of the field ${\bf m}$.
When we talk about correlation in the defect sector
we mean averages like in Eq.(\ref{eq:15a}) where there
is a vortex locating $\delta$-function inside the average.

By taking functional derivatives we are able to generate
the correlations between fields ${\bf m}$ at
arbitrary space-time points with the field at 
the space-time point $1$.
If we define
\be
Z_{H}(1)=\langle S(H)\delta (1)\rangle
\ee
then
\be
\langle \delta (1)m_{\nu_{2}}(2)m_{\nu_{3}}(3)
\ldots \rangle=Z_{H}^{-1}(1)
\frac{\delta}{\delta H_{\nu_{2}}(2)}
\frac{\delta}{\delta H_{\nu_{3}}(3)}\ldots Z_{H}(1)
~~~.
\ee
In our development a key property of the underlying
Gaussian distribution function is
\be
\langle m_{\nu_{1}}(1)A\rangle
=\sum_{\nu_{1'}}\int ~d\bar{1}G_{\nu_{1}\nu_{1'}}(1\bar{1})
\langle \frac{\delta}{\delta m_{\nu_{1'}}(\bar{1})}
A\rangle
\nonumber
\ee
for arbitrary $A$.  For $A=m_{\nu_{2}}(2)$
we obtain
\be
\langle m_{\nu_{1}}(1)m_{\nu_{2}}(2)\rangle
=G_{\nu_{1}\nu_{2}}(12)
~~~.
\label{eq:15}
\ee
If we assume that the system is isotropic in
the vector space, then
$G_{\nu_{1}\nu_{2}}(12)=\delta_{\nu_{1}\nu_{2}}G(12)$
and
\be
\langle m_{\nu_{1}}(1)A\rangle
=\int ~d\bar{1}G(1\bar{1})
\langle \frac{\delta}{\delta m_{\nu_{1}}(\bar{1})}
A\rangle
~~~.
\label{eq:17}
\ee
We then need to work out
\be
\langle S(H)\rho (1)\rangle
=\langle S(H)\delta (1)
\frac{1}{n!}\epsilon_{\mu_{1},\mu_{2},...,\mu_{n}}
\epsilon_{\nu_{1},\nu_{2},...,\nu_{n}}
\nabla_{\mu_{1}}m_{\nu_{1}}
\nabla_{\mu_{2}}m_{\nu_{2}}....
\nabla_{\mu_{n}}m_{\nu_{n}}\rangle
~~~.
\ee
Using Eq.(\ref{eq:17}) we have
\be
\langle S(H)\rho (1)\rangle
=\frac{1}{n!}\epsilon_{\mu_{1},\mu_{2},...,\mu_{n}}
\epsilon_{\nu_{1},\nu_{2},...,\nu_{n}}
\nabla_{\mu_{1}}G(1\bar{1})
\langle \frac{\delta}{\delta m_{\nu_{1}}(\bar{1})}
\left(S(H)\delta (1)
\nabla_{\mu_{2}}m_{\nu_{2}}....
\nabla_{\mu_{n}}m_{\nu_{n}}\right)\rangle
~~~.
\ee
The derivative of $S(H)$ leads to the introduction
of the quantity
\be
A_{\nu_{1}}(1)=\int d\bar{1}G(1\bar{1})H_{\nu_{1}}(\bar{1})
~~~.
\ee
We assume, and check self-consistently, that
$\left(\nabla_{\mu_{1}}G(1\bar{1})\right)|_{1=\bar{1}}
=0$.
The derivatives of the product
$\nabla_{\mu_{2}}m_{\nu_{2}}....
\nabla_{\mu_{n}}m_{\nu_{n}}$ with respect to 
$m_{\nu_{1}}(\bar{1})$ lead to contributions which
all vanish because it picks out terms
$\delta_{\nu_{1}\nu_{j}}$ which multiplies
$\epsilon_{\nu_{1},\nu_{2},.,\nu_{j},.,\nu_{n}}$ and 
$\epsilon_{\nu_{1},\nu_{2},.,\nu_{1},.,\nu_{n}}
=0$.
We have then 
\be
\langle S(H)\rho (1)\rangle
=\frac{1}{n!}\epsilon_{\mu_{1},\mu_{2},...,\mu_{n}}
\epsilon_{\nu_{1},\nu_{2},...,\nu_{n}}
\nabla_{\mu_{1}}A_{\nu_{1}}(1)
\langle S(H)\delta (1)
\nabla_{\mu_{2}}m_{\nu_{2}}....
\nabla_{\mu_{n}}m_{\nu_{n}}\rangle
~~~.
\ee
Clearly we can go through this process $n-1$ more 
times to obtain
$\langle S(H)\rho (1)\rangle={\cal D}_{A}(1)Z_{H}(1)$
where
\be
{\cal D}_{A}(1)
=\frac{1}{n!}\epsilon_{\mu_{1},\mu_{2},...,\mu_{n}}
\epsilon_{\nu_{1},\nu_{2},...,\nu_{n}}
\nabla_{\mu_{1}}A_{\nu_{1}}(1)
\nabla_{\mu_{2}}A_{\nu_{2}}(1)
....
\nabla_{\mu_{n}}A_{\nu_{n}}(1)
~~~.
\ee

Next we look at the current contributions to Eq.(\ref{eq:8})
in the form
\be
\langle S(H)\vec{\nabla}\cdot
\left(\rho (1) {\bf v} (1)\right)\rangle
=-\nabla_{\mu_{1}}^{(1)}
\frac{1}{(n-1)!}\epsilon_{\mu_{1},\mu_{2},...,\mu_{n}}
\epsilon_{\nu_{1},\nu_{2},...,\nu_{n}}
\langle S(H)\delta (1)
\dot{m}_{\nu_{1}}
\nabla_{\mu_{2}}m_{\nu_{2}}....
\nabla_{\mu_{n}}m_{\nu_{n}}\rangle
~~~.
\label{eq:33}
\ee
We assume that 
\be
\delta (1)\dot{m}_{\nu_{1}}(1)=-\delta (1)
\hat{O}(1)m_{\nu_{1}}(1)
\label{eq:34}
\ee
where $\hat{O}(1)$ is a derivative operator
defined by Eq.(\ref{eq:5}). In
ref.\onlinecite{mazVV} one has the choice
$\hat{O}(1)=-\nabla^{2}_{1}$.  After inserting
Eq.(\ref{eq:34}) back into Eq.(\ref{eq:33}) and
using Eq.(\ref{eq:17}) we
obtain
\be
\langle S(H)\vec{\nabla}
\cdot \left(\rho (1) {\bf v} (1)\right)\rangle
=-\nabla_{\mu_{1}}^{(1)}
\frac{1}{(n-1)!}\epsilon_{\mu_{1},\mu_{2},...,\mu_{n}}
\epsilon_{\nu_{1},\nu_{2},...,\nu_{n}}
\nonumber
\ee
\be
\times\Bigg[
\left(-\hat{O}(1)A_{\nu_{1}}(1)\right)
\nabla_{\mu_{2}}A_{\nu_{2}}(1)
....
\nabla_{\mu_{n}}A_{\nu_{n}}(1)Z_{H}(1)
\nonumber
\ee
\be
+\left(-\hat{O}(1)G(1\bar{1})\right)_{1=\bar{1}}
\langle S(H)\delta_{\nu_{1}}(1)
\nabla_{\mu_{2}}m_{\nu_{2}}....
\nabla_{\mu_{n}}m_{\nu_{n}}\rangle
\Bigg]
\ee
where
$\delta_{\nu_{1}}(1)=
\frac{\partial}{\partial m_{\nu_{1}}(1)}
\delta ({\bf m} (1))$
and the derivatives of the product of
$m$'s vanish as in the previous case.
Clearly the reduction of the term containing
$\delta_{\nu_{1}}(1)$ follows just as for 
$\langle S(H)\rho (1)\rangle $ with the result:
\be
\langle S(H)\vec{\nabla}\cdot
\left(\rho (1) {\bf v} (1)\right)\rangle
=\nabla_{\mu_{1}}^{(1)}
\frac{1}{(n-1)!}\epsilon_{\mu_{1},\mu_{2},...,\mu_{n}}
\epsilon_{\nu_{1},\nu_{2},...,\nu_{n}}
\nonumber
\ee
\be
\times\Bigg[
\left(\hat{O}(1)A_{\nu_{1}}(1)\right)
\nabla_{\mu_{2}}A_{\nu_{2}}(1)
....
\nabla_{\mu_{n}}A_{\nu_{n}}(1)Z_{H}(1)
\nonumber
\ee
\be
+\left(\hat{O}(1)G(1\bar{1})\right)_{1=\bar{1}}
\nabla_{\mu_{2}}A_{\nu_{2}}(1)
....
\nabla_{\mu_{n}}A_{\nu_{n}}(1)
\langle S(H)\delta_{\nu_{1}}(1)\rangle
\Bigg]
~~~.
\label{eq:40}
\ee

We must next evaluate $Z_{H}(1)$ and the
related quantity
$\langle S(H)\delta_{\nu_{1}}(1)\rangle$.
Determination of $Z_{H}(1)$ involves evaluation of the
Gaussian average
\be
Z_{H}(1)=\langle \delta ({\bf m}(1))S({\bf H})\rangle
=\int~\frac{d^{d}k}{(2\pi )^{d}}
\langle e^{i{\bf k}\cdot{\bf m}(1)}
e^{{\bf H}(\bar{1})\cdot{\bf m}(\bar{1})}
\rangle
~~~,
\ee
with the result
\be
Z_{H}(1)=\frac{e^{-\frac{1}{2}\frac{A^{2}(1)}{S_{0}(1)}}}
{\left(2\pi S_{0}(1)\right)^{n/2}}
exp\left[\frac{1}{2}H_{\nu_{1}}(\bar{1})H_{\nu_{1}}(\bar{2})
G(\bar{1}\bar{2})\right]
~~~.
\label{eq:40a}
\ee
where $S_{0}(1)=G(11)$.
Next we need 
\be
\langle S(H)\delta_{\nu_{1}}(1)\rangle
=\int~\frac{d^{d}k}{(2\pi )^{d}}ik_{\nu_{1}}
\langle e^{i{\bf k}\cdot{\bf m}(1)}
e^{{\bf H}(\bar{1})\cdot{\bf m}(\bar{1})}
\rangle
\nonumber
\ee
\be
=\int~\frac{d^{d}k}{(2\pi )^{d}}ik_{\nu_{1}}
exp\left[-\frac{1}{2}k^{2}G(11)+i{\bf k}\cdot{\bf A}(1)
+\frac{1}{2}H_{\nu_{2}}(\bar{1})H_{\nu_{2}}(\bar{2})
G(\bar{1}\bar{2})\right]
\nonumber
\ee
\be
=exp\left[\frac{1}{2}H_{\nu_{1}}(\bar{1})H_{\nu_{1}}(\bar{2})
G(\bar{1}\bar{2})\right]
\frac{\partial}{\partial A_{\nu_{1}}(1)}
exp\left[-\frac{1}{2}H_{\nu_{1}}(\bar{1})H_{\nu_{1}}(\bar{2})
G(\bar{1}\bar{2})\right]
Z{_H}(1)
\nonumber
\ee
\be
=-\frac{A_{\nu_{1}}(1)}{S_{0}(1)}Z_{H}(1)
~~~.
\label{eq:52}
\ee
Putting Eqs.(\ref{eq:40a}) and (\ref{eq:52}) back 
into Eq.(\ref{eq:40}), and
allowing the gradient $\nabla_{\mu_{1}}^{(1)}$ to act 
gives
\be
\langle S(H)\vec{\nabla}\cdot
\left(\rho (1) {\bf v} (1)\right)\rangle
=-{\cal D}_{B}(1)Z_{H}(1)
-{\cal D}_{\mu_{1}}^{B}(1)\nabla_{\mu_{1}}^{(1)}Z_{H}(1)
\ee
where
\be
{\cal D}_{B}(1)=\frac{1}{(n-1)!}\epsilon_{\mu_{1},\mu_{2},...,\mu_{n}}
\epsilon_{\nu_{1},\nu_{2},...,\nu_{n}}
\nabla_{\mu_{1}}B_{\nu_{1}}(1)
\nabla_{\mu_{2}}A_{\nu_{2}}(1)
....
\nabla_{\mu_{n}}A_{\nu_{n}}(1)
\ee
and
\be
{\cal D}_{\mu_{1}}^{B}(1)
=\frac{1}{(n-1)!}\epsilon_{\mu_{1},\mu_{2},...,\mu_{n}}
\epsilon_{\nu_{1},\nu_{2},...,\nu_{n}}
B_{\nu_{1}}(1)
\nabla_{\mu_{2}}A_{\nu_{2}}(1)
....
\nabla_{\mu_{n}}A_{\nu_{n}}(1)
~~~.
\ee
\be
B_{\nu_{1}}(1)=\left(-\hat{O}(1)+\Omega (1)\right)
A_{\nu_{1}}(1)
\ee
and
\be
\Omega (1)=\frac{1}{S_{0}(1)}
\left(\hat{O} (1)G(12)\right)_{1=2}
~~~.
\label{eq:38}
\ee
Putting the results together in Eq.(\ref{eq:8})
we obtain
\be
\frac{\partial}{\partial t_{1}}
\left({\cal D}_{A}(1)Z_{H}(1)\right)
={\cal D}_{B}(1)Z_{H}(1)
+{\cal D}_{\mu_{1}}^{B}(1)\nabla_{\mu_{1}}^{(1)}Z_{H}(1)
~~~.
\ee
We can write this in the form:
\be
\frac{\partial {\cal D}_{A}(1)}{\partial t_{1}}
+{\cal D}_{A}(1)\frac{\partial}{\partial t_{1}}
ln~Z_{H}(1)
={\cal D}_{B}(1)
+{\cal D}_{\mu_{1}}^{B}(1)\nabla_{\mu_{1}}
ln~Z_{H}(1)
~~~.
\label{eq:61}
\ee
After taking the derivatives $Z_{H}(1)$,
we can then write Eq.(\ref{eq:61}) in the form
\be
W_{2}(1)=W_{4}(1)
\label{eq:65}
\ee
where
\be
W_{2}(1)=\frac{\partial {\cal D}_{A}(1)}{\partial t_{1}}
-{\cal D}_{A}(1)\frac{n}{2}\frac{\dot{S}_{0}(1)}{S_{0}(1)}
-{\cal D}_{B}(1)
\ee
and
\be
W_{4}(1)={\cal D}_{A}(1)
\left( \frac{{\bf A}(1)}{S_{0}(1)}\cdot\dot{\bf A}(1)
-\frac{1}{2}A^{2}(1)\frac{\dot{S}_{0}(1)}{S_{0}^{2}(1)}\right)
-{\cal D}_{\mu_{1}}^{B}(1)\frac{{\bf A}(1)}{S_{0}(1)}\cdot
\nabla_{\mu_{1}}{\bf A}(1)
~~~.
\ee
Look first at $W_{2}(1)$ which can be written in the
form:
\be
W_{2}(1)=
\frac{1}{(n-1)!}\epsilon_{\mu_{1},\mu_{2},...,\mu_{n}}
\epsilon_{\nu_{1},\nu_{2},...,\nu_{n}}
\nonumber
\ee
\be
\times
\nabla_{\mu_{1}}\left(\dot{A}_{\nu_{1}}(1)
-\frac{1}{2}\frac{\dot{S}_{0}(1)}{S_{0}(1)}A_{\nu_{1}}(1)
-B_{\nu_{1}}(1)\right)
\nabla_{\mu_{2}}A_{\nu_{2}}(1)
....
\nabla_{\mu_{n}}A_{\nu_{n}}(1)
\nonumber
\ee
\be
=\frac{1}{(n-1)!}\epsilon_{\mu_{1},\mu_{2},...,\mu_{n}}
\epsilon_{\nu_{1},\nu_{2},...,\nu_{n}}
\nabla_{\mu_{1}} g_{\nu_{1}}(1)\nabla_{\mu_{2}}A_{\nu_{2}}(1)
....
\nabla_{\mu_{n}}A_{\nu_{n}}(1)
\label{eq:57}
\ee
where
\be
g_{\nu_{1}}(1)=
\dot{A}_{\nu_{1}}(1)
-\frac{1}{2}\frac{\dot{S}_{0}(1)}{S_{0}(1)}A_{\nu_{1}}(1)
-B_{\nu_{1}}(1)
~~~.
\ee

In looking at $W_{4}$ we need to focus on the quantity
\be
{\cal D}_{\mu_{1}}^{B}(1) A_{\nu}(1)
\nabla_{\mu_{1}} A_{\nu}(1)
\nonumber
\ee
\be
=\frac{1}{(n-1)!}\epsilon_{\mu_{1},\mu_{2},...,\mu_{n}}
\epsilon_{\nu_{1},\nu_{2},...,\nu_{n}}
B_{\nu_{1}}(1)\nabla_{\mu_{2}}A_{\nu_{2}}(1)
....
\nabla_{\mu_{n}}A_{\nu_{n}}(1)
A_{\nu}(1)\nabla_{\mu_{1}} A_{\nu}(1)
~~~.
\label{eq:73}
\ee
Note that
\be
\epsilon_{\mu_{1},\mu_{2},...,\mu_{n}}
\nabla_{\mu_{1}} A_{\nu}(1)\nabla_{\mu_{2}}A_{\nu_{2}}(1)
....
\nabla_{\mu_{n}}A_{\nu_{n}}(1)
=\epsilon_{\nu,\nu_{2},...,\nu_{n}}{\cal D}_{A}(1)
~~~.
\ee
Putting this back into Eq.(\ref{eq:73}), we find
\be
{\cal D}_{\mu_{1}}^{B}(1) A_{\nu}(1)
\nabla_{\mu_{1}} A_{\nu}(1)
=\frac{1}{(n-1)!}
\epsilon_{\nu_{1},\nu_{2},...,\nu_{n}}
\epsilon_{\nu,\nu_{2},...,\nu_{n}}
B_{\nu_{1}}(1)A_{\nu}(1){\cal D}_{A}(1)
\nonumber
\ee
\be
=B_{\nu_{1}}(1)A_{\nu_{1}}(1){\cal D}_{A}(1)
\ee
and
\be
W_{4}(1)={\cal D}_{A}(1)
\frac{A_{\nu}(1)}{S_{0}(1)}
\left(\dot{A}_{\nu_{1}}(1)
-\frac{1}{2}\frac{\dot{S}_{0}(1)}{S_{0}(1)}A_{\nu_{1}}(1)
-B_{\nu_{1}}(1)\right)
\nonumber
\ee
\be
={\cal D}_{A}(1)\frac{A_{\nu}(1)}{S_{0}(1)}g_{\nu}(1)
~~~.
\label{eq:62}
\ee
Using Eqs.(\ref{eq:57}) and (\ref{eq:62})
in Eq.(\ref{eq:65}) we find a solution for
general ${\bf H}$ if
\be
g_{\nu}(1)=
\dot{A}_{\nu}(1)
-\frac{1}{2}\frac{\dot{S}_{0}(1)}{S_{0}(1)}A_{\nu}(1)
-B_{\nu}(1)=0
~~~.
\ee
This will hold for all source fields ${\bf H}$ if
\be
\frac{\partial}{\partial t_{1}}G(12)
-\frac{1}{2}\frac{\dot{S}_{0}(1)}{S_{0}(1)}G(12)
=\left(-\hat{O}(1)+\Omega (1)\right)G(12)
~~~.
\label{eq:80}
\ee
Thus the average of the continuity equation, 
Eq.(\ref{eq:8}), is satisfied by a Gaussian
probability distribution if the associated
variance, $G(12)$, satisfies Eq.(\ref{eq:80}).

\section{Solution for $G(12)$}

We can solve Eq.(\ref{eq:80}) for $G(12)$ in some generality.  The
first step is to write
\be
G(12)=\sqrt{S_{0}(1)S_{0}(2)}f(12)
\label{eq:81}
\ee
where $S_{0}(1)=G(11)$.
Inserting this form into Eq.(\ref{eq:80}) gives
\be
\frac{\partial}{\partial t_{1}}f(12)
=\left(-\hat{O}(1)+\Omega (1)\right)f(12)
~~~.
\ee
We assume that the system is translationally invariant
and on Fourier transformation the operator $\hat{O}(1)$
is {\it diagonalized} and time independent.  We have
then
\be
\frac{\partial}{\partial t_{1}}f(q,t_{1},t_{2})
=\left(-O(q)+\Omega (1)\right)f(q,t_{1},t_{2})
~~~.
\label{eq:83}
\ee
We see that $\Omega (1)$, defined by Eq.(\ref{eq:38}),
is determined by the constraint
\be
\Omega (1)=\int~\frac{d^{d}q}{(2\pi )^{d}}
O(q)f(q,t_{1},t_{1})
~~~.
\label{eq:84}
\ee
We also have the equation
\be
\frac{\partial}{\partial t_{2}}f(q,t_{1},t_{2})
=\left(-O(q)+\Omega (2)\right)f(q,t_{1},t_{2})
~~~.
\label{eq:85}
\ee
Adding Eqs.(\ref{eq:83}) and (\ref{eq:85})
and setting $t_{1}=t_{2}=t$ we obtain for the
equal-time correlation function $f(q,t)\equiv f(q,t,t)$:
\be
\frac{\partial}{\partial t}f(q,t)
=2\left(-O(q)+\Omega (2)\right)f(q,t)
~~~.
\label{eq:85a}
\ee
For equal times we have from Eq.(\ref{eq:81}) that
$f(11)=1$ or
\be
1=\int~\frac{d^{d}q}{(2\pi )^{d}}
f(q,t)
~~~.
\label{eq:76}
\ee
The partial solution for Eq.(\ref{eq:85a}) is given by
\be
f(q,t)=exp\left(2\int_{t_{0}}^{t}d\tau
\left(\Omega (\tau )-O(q)\right)\right)f(q,t_{0})
\nonumber
\ee
\be
=R^{2}(t,t_{0})e^{-2O(q)(t-t_{0})}f(q,t_{0})
\label{eq:87}
\ee
where
\be
R(t_{1},t_{2})=exp\left(\int_{t_{2}}^{t_{1}}d\tau
~\Omega (\tau )\right)
~~~.
\label{eq:88}
\ee

We then need to solve for $\Omega (t)$.
Inserting Eq.(\ref{eq:87}) into Eq.(\ref{eq:76})
gives
\be
1=R^{2}(t,t_{0})I(t,t_{0})
\label{eq:70}
\ee
where
\be
I(t,t_{0})=\int~\frac{d^{d}q}{(2\pi )^{d}}
e^{-2O(q)(t-t_{0})}f(q,t_{0})
~~~.
\label{eq:75}
\ee
We then have from Eq.(\ref{eq:70})
\be
R^{2}(t,t_{0})=I^{-1}(t,t_{0})
~~~.
\ee
The constraint condition, Eq.(\ref{eq:84}), is given by 
\be
\Omega (t)=R^{2}(t,t_{0})\int~\frac{d^{d}q}{(2\pi )^{d}}
O(q)e^{-2O(q)(t-t_{0})}f(q,t_{0})
\nonumber
\ee
\be
=-\frac{1}{2}\frac{\dot{I}(t,t_{0})}{I(t,t_{0})}
~~~.
\ee
Thus the determination of $\Omega (1)$ is reduced to
evaluation of the integral $I(t,t_{0})$.  The equal time correlation
function is given then by
\be
f(q,t)=I^{-1}(t,t_{0})
e^{-2O(q)(t-t_{0})}f(q,t_{0})
~~~.
\label{eq:84a}
\ee
Going back to the unequal time correlation function
we can integrate Eq.({\ref{eq:83}).
\be
f(q,t_{1},t_{2})=exp\left(\int_{t_{2}}^{t_{1}}d\tau
\left(\Omega (\tau )-O(q)\right)\right)f(q,t_{2})
\nonumber
\ee
\be
=R(t_{1},t_{2})e^{-O(q)(t_{1}-t_{2})}f(q,t_{2})
\nonumber
\ee
\be
=R(t_{1},t_{0})R(t_{2},t_{0})e^{-O(q)(t_{1}+t_{2}-2t_{0})}
f(q,t_{0})
\ee
where we have used
$R(t_{1},t_{2})=R(t_{1},t_{0})/R(t_{2},t_{0})$.
We obtain a complete solution once one does
the integral for $I(t,t_{0})$.  Notice that
these results are independent of the specific form
for $S_{0}(t)$.

\section{Evaluation of Vortex Velocity Probability
Distribution}

The vortex velocity
probability distribution function defined by
\be
n_{0} P({\bf V},t)\equiv
\langle n(1)\delta ({\bf V}-{\bf v}(1))\rangle
\ee
where ${\bf v}$ as a function of the order
parameter is given by Eq.(\ref{eq:10}) with
$\vec{\psi}$ replaced by $\vec{m}$,
$n(1) =\delta ({\bf m}(1))|{\cal D}({\bf m}(1))|$ is the 
unsigned defect density,
and $n_{0}(1)=\langle n(1)\rangle$.
We notice in evaluating $P({\bf V},t)$ that it is of the
{\it defect~sector~form},  thus there is a defect locating
$\delta$-function in the average via the factor of $n(1)$.
Our results from section 2 suggest that in this sector
we can treat the field ${\bf m}$ as Gaussian with
variance given by $G(12)$ calculated in section 3.

In carrying out the average we need the auxiliary quantity
\be
W(\xi ,{\bf b})=\langle \delta ({\bf m} )
\prod_{\mu ,\nu}\delta (\xi_{\mu}^{\nu}-\nabla_{\mu}m_{\nu })
\delta ({\bf b}-{\bf K}\rangle
\ee
where
${\bf K}(1)=\hat{O}(1){\bf m} (1)$.
Then
\be
n_{0}P({\bf V},t)=
\int d^{n}b \prod_{\mu ,\nu}d\xi_{\mu}^{\nu}
|{\cal D}(\xi )|
\delta ({\bf V}-{\bf v}({\bf b},\xi ))W(\xi ,{\bf b})
\label{eq:105}
\ee
where
${\bf v}({\bf b},\xi ))
={\bf J}({\bf b},\xi)/ {\cal D}(\xi )$
with
\be
{\cal D}(\xi )=\frac{1}{n!}\epsilon_{\mu_{1},\mu_{2},...,\mu_{n}}
\epsilon_{\nu_{1},\nu_{2},...,\nu_{n}}
\xi_{\mu_{1}}^{\nu_{1}}
\xi_{\mu_{2}}^{\nu_{2}}....
\xi_{\mu_{n}}^{\nu_{n}}
\ee
and
\be
J_{\alpha}({\bf b},\xi )
=\frac{1}{(n-1)!}\epsilon_{\alpha,\mu_{2},...,\mu_{n}}
\epsilon_{\nu_{1},\nu_{2},...,\nu_{n}}
b_{\nu_{1}}
\xi_{\mu_{2}}^{\nu_{2}}....
\xi_{\mu_{n}}^{\nu_{n}} ~~~.
\ee

Let us turn to the Gaussian average determining
$W(\xi ,{\bf b})$.  Following the analysis used to
evaluate $Z_{H}$ in section 2 we introduce the
integral representation for the $\delta$-function
to obtain:
\be
W(\xi ,{\bf b})=\int ~\frac{d^{n}k}{(2\pi )^{n}}
\frac{d^{n}q}{(2\pi )^{n}}
\left(\prod_{\mu ,\nu}\int \frac{ds_{\mu}^{\nu}}{2\pi}\right)
e^{-i{\bf b}\cdot{\bf q}}
e^{-i\xi_{\mu}^{\nu}s_{\mu}^{\nu}}
\Gamma ({\bf k},{\bf q},s)
\label{eq:107}
\ee
where
\be
\Gamma ({\bf k},{\bf q},s)
=\langle e^{i{\bf k}\cdot{\bf m}(1)}
e^{i{\bf q}\cdot{\bf K}(1)}
e^{is_{\mu}^{\nu}\nabla_{\mu}m_{\nu}(1)}
\rangle
~~~.
\ee
If we introduce
\be
H_{\alpha}(\bar{1})=
i\left[k_{\alpha}+q_{\alpha}\hat{O}(1)
+s_{\mu}^{\alpha}\nabla_{\mu}^{(1)}\right]
\delta(\bar{1}1)
\ee
then we can write
\be
\Gamma ({\bf k},{\bf q},s)
=\langle e^{H_{\alpha}(\bar{1})m_{\alpha}(\bar{1})}\rangle
=exp\left[\frac{1}{2}H_{\alpha}(\bar{1})
H_{\alpha}(\bar{2})G(\bar{1}\bar{2})\right]
\label{eq:86}
\ee
where $G(12)$ was determined in section 3.  
We assume that the cross terms involving an odd
number of gradients vanishes in the argument of the
exponential.  We have then
\be
\Gamma ({\bf k},{\bf q},s)
=exp\left[-\frac{1}{2}\left(
k^{2}S_{0}(1)+2{\bf k}\cdot{\bf q}S_{c}(1)
+q^{2}S_{O2}(1)+s_{\mu}^{\alpha}s_{\mu}^{\alpha}
S_{\mu}^{(2)}(1)\right)
\right]
\ee
where
\be
\frac{S_{c}(1)}{S_{0}(1)}=\int~\frac{d^{n}q}{(2\pi )^{n}}O(q)f(q,t)
=\Omega (1)=-\frac{1}{2}\frac{\dot{I}(1)}{I(1)}
~~~,
\label{eq:100}
\ee
where we have used Eq.(\ref{eq:83}),
\be
\frac{S_{O2}(1)}{S_{0}(1)}=\int~\frac{d^{n}q}{(2\pi )^{n}}O^{2}(q)f(q,t)
=\frac{1}{4}\frac{\ddot{I}(1)}{I(1)}
\label{eq:101}
\ee
having used Eq.(\ref{eq:84a}), and
\be
\frac{S_{\mu\mu'}^{(2)}(1)}{S_{0}(1)}=\int~\frac{d^{n}q}{(2\pi )^{n}}
q_{\mu }q_{\mu '}f(q,t_{1})
=\delta_{\mu\mu '}\frac{S_{\mu}^{(2)}(1)}{S_{0}(1)}
~~~.
\label{eq:102}
\ee

The next step in extracting $W$ is to integrate over
${\bf k}$:  
\be
\Gamma ({\bf q}, s)=\int ~\frac{d^{n}k}{(2\pi )^{n}}
\Gamma ({\bf k},{\bf q}, s)
=\frac{1}{(2\pi S_{0})^{n/2}}
e^{-\frac{1}{2}q^{2}\bar{S}}
e^{-\frac{1}{2}s_{\mu}^{\alpha}s_{\mu}^{\alpha}
S_{\mu}^{(2)}}
\label{eq:123}
\ee
where
\be
\bar{S}=S_{O2}-\frac{S_{c}^{2}}{S_{0}}
~~~.
\label{eq:108}
\ee
Using Eq.(\ref{eq:123}) back in Eq.(\ref{eq:107})
we obtain
\be
W(\xi ,{\bf b})=\int 
\frac{d^{n}q}{(2\pi )^{n}}
\left(\prod_{\mu ,\nu}\int \frac{ds_{\mu}^{\nu}}{2\pi}\right)
e^{-i{\bf b}\cdot{\bf q}}
e^{-i\xi_{\mu}^{\nu}s_{\mu}^{\nu}}
\Gamma ({\bf q},s)
\nonumber
\ee
\be
=\int 
\frac{d^{n}q}{(2\pi )^{n}}
\left(\prod_{\mu ,\nu}\int \frac{ds_{\mu}^{\nu}}{2\pi}\right)
e^{-i{\bf b}\cdot{\bf q}}
e^{-i\xi_{\mu}^{\nu}s_{\mu}^{\nu}}
\frac{1}{(2\pi S_{0})^{n/2}}
e^{-\frac{1}{2}q^{2}\bar{S}}
e^{-\frac{1}{2}s_{\mu}^{\alpha}s_{\mu}^{\alpha}
S_{\mu}^{(2)}}
~~~.
\ee
This factorizes into a product of three natural parts
\be
W(\xi ,{\bf b})=\frac{1}{(2\pi S_{0})^{n/2}}
W(\xi)W({\bf b})
\ee
where
\be
W(\xi)=\left(\prod_{\mu ,\nu} 
\int ~\frac{ds_{\mu}^{\nu}}{2\pi}\right)
e^{-i\xi_{\mu}^{\nu}s_{\mu}^{\nu}}
e^{-\frac{1}{2}s_{\mu}^{\alpha}s_{\mu}^{\alpha}
S_{\mu}^{(2)}}
\ee
and
\be
W({\bf b})=\int
\frac{d^{n}q}{(2\pi )^{n}}
e^{-i{\bf b}\cdot{\bf q}}e^{-\frac{1}{2}q^{2}\bar{S}}
~~~.
\ee
Using the basic integral
\be
\int\frac{dx}{2\pi}e^{-iyx}e^{-\frac{a}{2}x^{2}}
=\frac{1}{\sqrt{2\pi a}}e^{-\frac{y^{2}}{2a}}
\ee
we can evaluate both factors:
\be
W(\xi)=\left(\prod_{\mu}\frac{1}{2\pi S_{\mu}^{(2)}}\right)^{n/2}
e^{-\frac{(\xi_{\mu'}^{\nu})^{2}}{2S_{\mu'}^{(2)}}}
~~~,
\ee
\be
W({\bf b})=
\frac{1}{(2\pi \bar{S})^{n/2}}e^{-\frac{b^{2}}{2\bar{S}}}
~~~.
\ee

Turning to $n_{0}P({\bf V},t)$ given by Eq.(\ref{eq:105}),
we see that we have the integral over ${\bf b}$ of the
form
\be
J_{b}=\int ~ d^{n}b 
~\delta ({\bf V}-{\bf v}({\bf b},\xi ))W({\bf b})
\nonumber
\ee
\be
=\int ~ d^{n}b \int ~\frac{d^{n}z}{(2\pi )^{n}}
e^{-i{\bf V}\cdot{\bf z}}
e^{i{\bf v}({\bf b},\xi )\cdot{\bf z}}
W({\bf b})
~~~.
\ee
We can then write
${\bf z}\cdot{\bf v}({\bf b},\xi )={\bf a}\cdot{\bf b}$
where
\be
a_{\nu_{1}}=
\frac{1}{{\cal D}(n-1)!}z_{\alpha}
\epsilon_{\alpha,\mu_{2},...,\mu_{n}}
\epsilon_{\nu_{1},\nu_{2},...,\nu_{n}}
\xi_{\mu_{2}}^{\nu_{2}}....
\xi_{\mu_{n}}^{\nu_{n}} 
\ee
then
\be
J_{b}=
\int ~ d^{n}b \int ~\frac{d^{n}z}{(2\pi )^{n}}
e^{-i{\bf V}\cdot{\bf z}}
e^{i{\bf a}\cdot{\bf b}}
\frac{1}{(2\pi \bar{S})^{n/2}}e^{-\frac{b^{2}}{2\bar{S}}}
\nonumber
\ee
\be
=\int ~\frac{d^{n}z}{(2\pi )^{n}}
e^{-i{\bf V}\cdot{\bf z}}
e^{-\frac{\bar{S}}{2}{\bf a}^{2}}
\label{eq:139}
\ee
where
${\bf a}^{2}=z_{\alpha}M_{\alpha\beta}z_{\beta}$
and the matrix $M$ is given by
\be
M_{\alpha ,\beta }=\frac{1}{{\cal D}^{2}[(n-
1)!]^{2}}\epsilon_{\alpha,\mu_{2},...,\mu_{n}}
\epsilon_{\nu,\nu_{2},...,\nu_{n}}
\xi_{\mu_{2}}^{\nu_{2}}....\xi_{\mu_{n}}^{\nu_{n}}
\epsilon_{\beta,\mu_{2}',...,\mu_{n}'}
\epsilon_{\nu,\nu_{2}',...,\nu_{n}'}
\xi_{\mu_{2}'}^{\nu_{2}'}....\xi_{\mu_{n}'}^{\nu_{n}'}~~~~.
~~~.
\ee
Doing the remaining Gaussian $z$ integration in
Eq.(\ref{eq:139}) we obtain
\be
J_{b}=
\frac{1}{(2\pi\bar{S})^{n/2}}\frac{1}{\sqrt{det M}}
exp \Biggl[-\frac{1}{2\bar{S} }\sum_{\mu ,\nu}
V^{\mu}[M^{-1}]_{\mu ,\nu }V^{\nu}\Biggr]
\ee
and
\be
n_{0}P({\bf V},t)=\int \prod_{\mu ,\nu}d\xi_{\mu}^{\nu}
|{\cal D}(\xi )|W(\xi )
\frac{1}{(4\pi^{2}S_{0}\bar{S} )^{n/2}}\frac{1}{\sqrt{det M}}
exp \Biggl[-\frac{1}{2\bar{S}}\sum_{\mu ,\nu}
V^{\mu}[M^{-1}]_{\mu ,\nu }V^{\nu}\Biggr]
\ee

We must look at the matrix $M$ and its inverse.
First multiply $M_{\alpha\beta}$ by $\xi_{\alpha}^{\nu}$
to obtain
\be
\xi_{\alpha}^{\nu}M_{\alpha\beta}
=\xi_{\alpha}^{\nu}\frac{1}{{\cal D}^{2}[(n-
1)!]^{2}}\epsilon_{\alpha,\mu_{2},...,\mu_{n}}
\epsilon_{\nu_{1},\nu_{2},...,\nu_{n}}
\xi_{\mu_{2}}^{\nu_{2}}....\xi_{\mu_{n}}^{\nu_{n}}
\epsilon_{\beta,\mu_{2}',...,\mu_{n}'}
\epsilon_{\nu_{1},\nu_{2}',...,\nu_{n}'}
\xi_{\mu_{2}'}^{\nu_{2}'}....\xi_{\mu_{n}'}^{\nu_{n}'}~~~~.
~~~.
\ee
However
\be
\xi_{\mu_{1}}^{\nu}\epsilon_{\mu_{1},\mu_{2},...,\mu_{n}}
\xi_{\mu_{2}}^{\nu_{2}}....\xi_{\mu_{n}}^{\nu_{n}}
={\cal D}(\xi )\epsilon_{\nu,\nu_{2},...,\nu_{n}}
\ee
then
\be
\xi_{\alpha}^{\nu}M_{\alpha\beta}
=\frac{1}{{\cal D}^{2}[(n-1)!]^{2}}
{\cal D}(\xi )\epsilon_{\nu,\nu_{2},...,\nu_{n}}
\epsilon_{\nu_{1},\nu_{2},...,\nu_{n}}
\epsilon_{\beta,\mu_{2}',...,\mu_{n}'}
\epsilon_{\nu_{1},\nu_{2}',...,\nu_{n}'}
\xi_{\mu_{2}'}^{\nu_{2}'}....\xi_{\mu_{n}'}^{\nu_{n}'}~~~~.
\ee
\be
=\frac{1}{{\cal D}(n-1)!}\epsilon_{\beta,\mu_{2}',...,\mu_{n}'}
\epsilon_{\nu,\nu_{2}',...,\nu_{n}'}
\xi_{\mu_{2}'}^{\nu_{2}'}....\xi_{\mu_{n}'}^{\nu_{n}'}
\label{eq:147}
\ee
where we have used
\be
\epsilon_{\nu,\nu_{2},...,\nu_{n}}
\epsilon_{\nu_{1},\nu_{2},...,\nu_{n}}
=\delta_{\nu ,\nu_{1}}(n-1)!
~~~.
\ee
Multiply Eq.(\ref{eq:147}) by $\xi_{\gamma}^{\nu}$
to obtain 
\be
\xi_{\gamma}^{\nu}\xi_{\alpha}^{\nu}M_{\alpha\beta}
=\frac{1}{{\cal D}(n-1)!}
\xi_{\gamma}^{\nu}
\epsilon_{\beta,\mu_{2}',...,\mu_{n}'}
\epsilon_{\nu ,\nu_{2}',...,\nu_{n}'}
\xi_{\mu_{2}'}^{\nu_{2}'}....\xi_{\mu_{n}'}^{\nu_{n}'}
\nonumber
\ee
\be
=\frac{1}{{\cal D}(n-1)!}
\epsilon_{\beta,\mu_{2}',...,\mu_{n}'}
{\cal D}(\xi )\epsilon_{\gamma ,\mu_{2}',...,\mu_{n}'}
=\delta_{\gamma\beta}
~~~.
\ee
Thus we have the beautiful result:
\be
\left(M^{-1}\right)_{\alpha\beta}
=\sum_{\nu}\xi_{\alpha}^{\nu}\xi_{\beta}^{\nu}
~~~.
\label{eq:152}
\ee
We need $det~ M =1/det~ M^{-1}$.  We have
\be
det M^{-1}=\frac{1}{n!}
\epsilon_{\alpha_{1},\alpha_{2},...,\alpha_{n}}
\epsilon_{\beta_{1},\beta_{2},...,\beta_{n}}
\xi_{\alpha_{1}}^{\nu_{1}}\xi_{\beta_{1}}^{\nu_{1}}
\xi_{\alpha_{2}}^{\nu_{2}}\xi_{\beta_{2}}^{\nu_{2}}
\cdots
\xi_{\alpha_{n}}^{\nu_{n}}\xi_{\beta_{n}}^{\nu_{n}}
\nonumber
\ee
\be
=\frac{1}{n!}
\epsilon_{\nu_{1},\nu_{2},...,\nu_{n}}{\cal D}(\xi )
\epsilon_{\nu_{1},\nu_{2},...,\nu_{n}}{\cal D}(\xi )
\nonumber
\ee
\be
=\frac{1}{n!}{\cal D}(\xi )^{2}n!
={\cal D}(\xi )^{2}
~~~.
\ee
Using the clean result
$det (M) =1/({\cal D})^{2}$,
and Eq.(\ref{eq:152}), we have
\be
n_{0}P({\bf V},t)
=\int\left(\prod_{\mu ,\nu}
\frac{d\xi_{\mu}^{\nu}}{\sqrt{(2\pi S_{\mu}^{(2)})}}\right)
\frac{{\cal D}^{2}(\xi)}{(4\pi^{2}S_{0} \bar{S} )^{n/2}}
e^{-\frac{1}{2}A(\xi )}
\ee
where
\be
A(\xi )=\sum_{\mu,\nu}\frac{1}{S_{\mu}^{(2)}}
(\xi_{\mu}^{\nu})^{2}
+\frac{1}{\bar{S} }\sum_{\alpha ,\beta ,\nu}
V^{\alpha}\xi_{\alpha}^{\nu}\xi_{\beta}^{\nu}V^{\beta}
~~~.
\ee
Next make the change of variables
$\xi_{\mu}^{\nu}=\sqrt{S_{\mu}^{(2)}}\tilde{\xi}_{\mu}^{\nu}$
to obtain, ${\cal D}(\xi)
=\left(\prod_{\mu}\sqrt{S_{\mu}^{(2)}}\right)
{\cal D}(\tilde{\xi})$,
\be
n_{0}P({\bf V},t)=
\prod_{\mu}\left(\frac{S_{\mu}^{(2)}}
{2\pi\sqrt{S_{0}\bar{S}}}\right)
\int\left(\prod_{\mu ,\nu}
\frac{d\tilde{\xi}_{\mu}^{\nu}}{\sqrt{(2\pi)}}\right)
{\cal D}^{2}(\tilde{\xi})
e^{-\frac{1}{2}A(\tilde{\xi} )}
\label{eq:162}
\ee
where
\be
A(\tilde{\xi} )=\sum_{\mu,\nu}
(\tilde{\xi}_{\mu}^{\nu})^{2}
+\sum_{\alpha ,\beta ,\nu}
\tilde{V}^{\alpha}\tilde{\xi}_{\alpha}^{\nu}
\tilde{\xi}_{\beta}^{\nu}\tilde{V}^{\beta}
\ee
and
\be
\tilde{V}^{\alpha}=\sqrt{\frac{S_{\alpha}^{(2)}}
{\bar{S}}}V^{\alpha}
~~~.
\label{eq:133}
\ee
Next we make the transformation from 
$\tilde{\xi}_{\alpha}^{\nu}$ to $\chi_{\alpha}^{\nu}$
via
\be
\tilde{\xi}_{\alpha}^{\nu}=N_{\alpha ,\beta }\chi_{\beta }^{\nu}
\ee
such that
\be
A(\tilde{\xi} )
=\sum_{\alpha ,\nu}(\chi_{\alpha }^{\nu})^{2}
=\sum_{\mu,\nu}
N_{\mu,\bar{\mu}_{1}}N_{\mu,\bar{\mu}_{2}}
\chi_{\bar{\mu}_{1} }^{\nu}
\chi_{\bar{\mu}_{2} }^{\nu}
+\sum_{\alpha ,\beta ,\nu}
\tilde{V}^{\alpha}N_{\alpha ,\bar{\mu}_{1}}
\chi_{\bar{\mu}_{1} }^{\nu}
N_{\beta ,\bar{\mu}_{2}}
\chi_{\bar{\mu}_{2} }^{\nu}
\tilde{V}^{\beta}
~~~.
\ee
This requires that $N$ satisfy
\be
N_{\mu,\mu_{1}}N_{\mu,\mu_{2}}
+\tilde{V}^{\alpha}N_{\alpha ,\mu_{1}}
\tilde{V}^{\beta}N_{\beta ,\mu_{2}}
=\delta_{\mu_{1},\mu_{2}}
~~~.
\label{eq:168}
\ee
This has a solution given by
\be
N_{\alpha \beta}=\delta_{\alpha \beta}
+[\frac{1}{\sqrt{1+\tilde{V}^{2}}}-1]
\hat{V}^{\alpha}\hat{V}^{\beta}
~~~.
\ee

We then need the Jacobian of the transformation
$d\xi_{\mu}^{\nu}\rightarrow d\chi_{\mu}^{\nu}$
and ${\cal D}(\tilde{\xi})$ evaluated in
terms of $\chi$.  Look first at ${\cal D}(\tilde{\xi})$.
\be
{\cal D}(\tilde{\xi})=\frac{1}{n!}
\epsilon_{\alpha_{1}\alpha_{2}\ldots\alpha_{n}}
\epsilon_{\nu_{1}\nu_{2}\ldots\nu_{n}}
\nonumber
\ee
\be
\times\left[\chi_{\alpha_{1}}^{\nu_{1}}+N_{1}
\tilde{V}^{\alpha_{1}}\chi_{\bar{\mu}_{1}}^{\nu_{1}}
\tilde{V}^{\bar{\mu}_{1}}\right]
\left[\chi_{\alpha_{2}}^{\nu_{2}}+N_{1}
\tilde{V}^{\alpha_{2}}\chi_{\bar{\mu}_{2}}^{\nu_{2}}
\tilde{V}^{\bar{\mu}_{2}}\right]
\ldots
\left[\chi_{\alpha_{n}}^{\nu_{n}}+N_{1}
\tilde{V}^{\alpha_{n}}\chi_{\bar{\mu}_{n}}^{\nu_{n}}
\tilde{V}^{\bar{\mu}_{n}}\right]
~~~.
\ee
If we multiply this out in powers of $\tilde{{\bf V}}$
we see that if we have more than one factor of
$\tilde{V}^{\alpha_{i}}$ then the contribution vanishes
due to antisymmetry, therefore
\be
{\cal D}(\tilde{\xi})={\cal D}(\chi)
+\frac{1}{n!}n
\epsilon_{\alpha_{1}\alpha_{2}\ldots\alpha_{n}}
\epsilon_{\nu_{1}\nu_{2}\ldots\nu_{n}}
N_{1}\tilde{V}^{\alpha_{1}}\chi_{\bar{\mu}_{1}}^{\nu_{1}}
\tilde{V}^{\bar{\mu}_{1}}\chi_{\alpha_{2}}^{\nu_{2}}
\ldots\chi_{\alpha_{n}}^{\nu_{n}}
\ee
which can be put in the form:
\be
{\cal D}(\tilde{\xi})=\frac{{\cal D}(\chi)}
{\sqrt{1+\tilde{V}^{2}}}
~~~.
\ee
Next we need the Jacobian
\be
J=\prod_{\nu}det
\left( \frac{\partial \tilde{\xi}_{\mu}^{\nu}}
{\partial \chi_{\mu'}^{\nu}}\right)
\ee
\be
=\left[det\left(\delta_{\mu ,\mu'}
+N_{1}\tilde{V}^{\mu}\tilde{V}^{\mu'}\right)\right]^{n}
=\left(J_{0}\right)^{n}
\ee
where
\be
J_{0}=det\left(\delta_{\mu ,\mu'}
+N_{1}\tilde{V}^{\mu}\tilde{V}^{\mu'}\right)
\ee
\be
=\frac{1}{n!}
\epsilon_{\alpha_{1}\alpha_{2}\ldots\alpha_{n}}
\epsilon_{\beta_{1}\beta_{2}\ldots\beta_{n}}
\left[\delta_{\alpha_{1} ,\beta_{1}}
+N_{1}\tilde{V}^{\alpha_{1}}\tilde{V}^{\beta_{1}}\right]
\left[\delta_{\alpha_{2} ,\beta_{2}}
+N_{1}\tilde{V}^{\alpha_{2}}\tilde{V}^{\beta_{2}}\right]
\ldots
\left[\delta_{\alpha_{3} ,\beta_{3}}
+N_{1}\tilde{V}^{\alpha_{3}}\tilde{V}^{\beta_{3}}\right]
~~~.
\ee
Again, expanding this out in powers of $\tilde{V}$, only
the first two terms contribute due to symmetry and
we have
\be
J_{0}=\frac{1}{n!}
\epsilon_{\alpha_{1}\alpha_{2}\ldots\alpha_{n}}^{2}
+nN_{1}\epsilon_{\alpha_{1}\alpha_{2}\ldots\alpha_{n}}
\epsilon_{\beta_{1}\alpha_{2}\ldots\alpha_{n}}
\tilde{V}^{\alpha_{1}}\tilde{V}^{\beta_{1}}
\nonumber
\ee
\be
=1+N_{1}\tilde{V}^{2}
=\frac{1}{\sqrt{1+\tilde{V}^{2}}}
~~~.
\ee

Going back to Eq.(\ref{eq:162}) we have
\be
n_{0}P[{\bf V},t]=
\prod_{\mu}\left(\frac{S_{\mu}^{(2)}}
{2\pi\sqrt{S_{0}\bar{S}}}\right)
\frac{1}{(1+\tilde{V}^{2})^{(n+2)/2}}J_{F}
\ee
where we have the final integral
\be
J_{F}=
\int\prod_{\mu ,\nu}
\frac{d\chi_{\mu}^{\nu}}{\sqrt{2\pi}}
{\cal D}^{2}(\chi)
e^{-\frac{1}{2}A(\chi )}
~~~.
\ee
We can evaluate $J_{F}$ directly.  The first step is to
write:
\be
J_{F}=\int\prod_{\mu ,\nu}
\frac{d\chi_{\mu}^{\nu}}{\sqrt{2\pi}}
\epsilon_{\mu_{1}\mu_{2}\ldots\mu_{n}}
\epsilon_{\mu_{1}'\mu_{2}'\ldots\mu_{n}'}
\chi_{\mu_{1}}^{(1)}\chi_{\mu_{2}}^{(2)}
\ldots\chi_{\mu_{n}}^{(n)}
\chi_{\mu_{1}'}^{(1)}\chi_{\mu_{2}'}^{(2)}
\ldots\chi_{\mu_{n}'}^{(n)}
e^{-\frac{1}{2}\sum_{\mu\nu}(\chi_{\mu}^{\nu})^{2}}
~~~.
\ee
This factorizes into a product of integrals for
fixed $\nu$
\be
J_{F}=\epsilon_{\mu_{1}\mu_{2}\ldots\mu_{n}}
\epsilon_{\mu_{1}'\mu_{2}'\ldots\mu_{n}'}
\int\prod_{\mu}\frac{d\chi_{\mu}^{(1)}}{\sqrt{2\pi}}
\chi_{\mu_{1}}^{(1)}\chi_{\mu_{1}'}^{(1)}
e^{-\frac{1}{2}\sum_{\mu}(\chi_{\mu}^{(1)})^{2}}
\int\prod_{\mu}\frac{d\chi_{\mu}^{(2)}}{\sqrt{2\pi}}
\chi_{\mu_{1}}^{(2)}\chi_{\mu_{1}'}^{(2)}
e^{-\frac{1}{2}\sum_{\mu}(\chi_{\mu}^{(2)})^{2}}
\ldots
\nonumber
\ee
\be
\times\int\prod_{\mu}\frac{d\chi_{\mu}^{(n)}}{\sqrt{2\pi}}
\chi_{\mu_{1}}^{(n)}\chi_{\mu_{1}'}^{(n)}
e^{-\frac{1}{2}\sum_{\mu}(\chi_{\mu}^{(n)})^{2}}
\nonumber
~~~.
\ee
Each integral in the product is equal to $1$ except
for those giving a $\delta$-function with unit
coefficient:

\be
J_{F}
=\epsilon_{\mu_{1}\mu_{2}\ldots\mu_{n}}
\epsilon_{\mu_{1}'\mu_{2}'\ldots\mu_{n}'}
\delta_{\mu_{1},\mu_{1}'}
\delta_{\mu_{2},\mu_{2}'}
\ldots
\delta_{\mu_{n},\mu_{n}'}
\nonumber
\ee
\be
=\epsilon_{\mu_{1}\mu_{2}\ldots\mu_{n}}^{2}=n!
~~~.
\ee
We have then

\be
n_{0}P[{\bf V},t]=n!
\prod_{\mu}\left(\frac{S_{\mu}^{(2)}}
{2\pi\sqrt{S_{0}\bar{S}}}\right)
\frac{1}{(1+\tilde{V}^{2})^{(n+2)/2}}
~~~.
\label{eq:198}
\ee

Since $P({\bf V},t)$ is normalized to one, we find on integration over
${\bf V}$, the result
\be
n_{0}=
\frac{n!}{2^{n/2}\Gamma(\frac{n}{2}+1)}
\prod_{\mu}\sqrt{\frac{S_{\mu}^{(2)}}{2\pi S_{0}}}
\ee
which agrees with previous results in the isotropic
limit.  Eliminating $n_{0}$ in Eq.(\ref{eq:198})
we obtain the final result for the vvpdf:
\be
P[{\bf V},t]=\frac{\Gamma(\frac{n}{2}+1)}{\pi^{n/2}}
\left(\prod_{\mu}\frac{1}{\bar{v}_{\mu}}\right)
\frac{1}{(1+\tilde{V}^{2})^{(n+2)/2}}
\label{eq:161}
\ee
where
$\tilde{V}_{\alpha}={V}_{\alpha}/\bar{v}_{\alpha}$,
\be
\bar{v}_{\mu}=\sqrt{\frac{\bar{S}}{S_{\mu}^{(2)}}}
\label{eq:163}
~~~,
\ee
using Eqs.(\ref{eq:100}), (\ref{eq:101}) and (\ref{eq:108})
we have 
\be
\bar{S}=S_{0}\frac{1}{4}\left[\frac{\ddot{I}}{I}
-\left(\frac{\dot{I}}{I}\right)^{2}\right] 
\label{eq:145}
~~~,
\ee
while Eq.(\ref{eq:102}) gives
\be
S_{\mu}^{(2)}=S_{0}\frac{1}{I}
\int \frac{d^{d}q}{(2\pi )^{d}}q_{\mu}^{2}
e^{-2O(q)(t-t_{0})}f({\bf q},t_{0})
\label{eq:170}
\ee
with $I$ defined by Eq.(\ref{eq:80}):
\be
I=\int \frac{d^{d}q}{(2\pi )^{d}}
e^{-2O(q)(t-t_{0})}f({\bf q},t_{0})
~~~.
\label{eq:171}
\ee
The needed input to determine the average vortex speed,
$\bar{v}_{\mu}$, is the function $O({\bf q})$,
the Fourier transform of the operator defined
by Eq.(\ref{eq:5}), and the initial condition
$f({\bf q},t_{0})$.
This result for $P[{\bf V},t]$ is the anisotropic generalization
of Eq.(\ref{eq:1}).  For the set of models included here the
velocity tail exponent is $(n+2)$ independent of direction.

\section{Anisotropic Case}

As a particularly simple example,
suppose that we have the choice in the governing Langevin
equation 
\be
\hat{O}(1)\psi_{\nu}(1) =-\sum_{\alpha}c_{\alpha}
\nabla_{\alpha}^{2}\psi_{\nu}(1)
~~~,
\ee
or, in terms of Fourier transforms,
\be
O({\bf q})=\sum_{\alpha}c_{\alpha}q_{\alpha}^{2}
~~~.
\ee
The associated vortex-velocity probability
distribution is given by 
Eq.(\ref{eq:161}) and we need to work out the average 
vortex speed given by Eq.(\ref{eq:163}).
Assuming the initial condition 
\be
f(q,0)=\left(\prod_{\alpha}(2\pi h_{\alpha})^{1/2}\right)
e^{-\frac{1}{2}h_{\mu}q_{\mu}^{2}}
~~~,
\ee
we have from  Eq.(\ref{eq:171})
\be
I(t)=\int~\frac{d^{n}q}{(2\pi )^{n}}
\left(\prod_{\alpha}(2\pi h_{\alpha})^{1/2}\right)
e^{-\frac{1}{2}h_{\mu}(t)q_{\mu}^{2}}
=\prod_{\alpha}\left(\frac{ h_{\alpha}}
{ h_{\alpha}(t)}\right)^{1/2}
\ee
where
\be
h_{\alpha}(t)=h_{\alpha}+4c_{\alpha}(t-t_{0})
~~~,
\ee
from Eq.(\ref{eq:145})
\be
\frac{\bar{S}}{S_{0}}=
2\sum_{\mu}\left(\frac{c_{\mu}}{h_{\mu}(t)}\right)^{2}
~~~,
\ee
and from Eq.(\ref{eq:170})
\be
\frac{S_{\mu}^{(2)}(1)}{S_{0}(1)}=
\int~\frac{d^{n}q}{(2\pi )^{n}}
q_{\mu }^{2}f(q,t_{1}) =\frac{1}{h_{\mu}(t)}
~~~.
\ee
Putting these results back into Eq.(\ref{eq:163})
we find the scaling 
velocity for a simple anisotropic system
is given by
\be
\bar{v}^{2}_{\mu}=2h_{\mu}(t)
\sum_{\alpha}\left(\frac{c_{\alpha}}{h_{\alpha}(t)}\right)^{2}
\ee
\be
=2(h_{\mu}+4c_{\mu}(t-t_{0}))
\sum_{\alpha}\left(\frac{c_{\alpha}}
{h_{\alpha}+4c_{\alpha}(t-t_{0})}\right)^{2}
~~~.
\ee
In the large time limit we have
\be
\bar{v}^{2}_{\mu}=\frac{d}{2}\frac{c_{\mu}}{t}
\ee
and the final form is a simple generalization of the
isotropic result.

\section{Conserved Order Parameter Case}

Let us look at the COP case where
$O(q)=-q^{2}+q^{4}$.
We choose the rather general initial condition
\be
f(q,0)=\left(\frac{h}{2\pi}\right)^{d/2}
e^{-\frac{h}{2}q^{2}}
~~~,
\ee
which satisfies the normalization condition given
by Eq.(\ref{eq:76}).  We then need to evaluate the
integral $I$ and the numerator in
Eq.(\ref{eq:170}):
\be
J=\int \frac{d^{d}q}{(2\pi )^{d}}\frac{q^{2}}{d}
e^{2t(q^{2}-q^{4})}
\left(\frac{h}{2\pi}\right)^{d/2}
e^{-\frac{h}{2}q^{2}}
\nonumber
\ee
\be
=-\frac{2}{d}h^{d/2}\frac{\partial}{\partial h}
(h^{-d/2}I)
~~~.
\label{eq:160}
\ee
We see that all of the ingredients contributing to
$\bar{v}^{2}$ can be expressed in terms of $I$ and 
it's derivatives.  We see that $I$ can be written
in the form:
\be
I=\tilde{I}_{0}h^{d/2}\int_{0}^{\infty} dq q^{d-1}
e^{2(q^{2}-q^{4})t-\frac{h}{2}q^{2}}
\ee
where $\tilde{I}_{0}$ is a constant which depends on
$d$ and cancels when we take ratios.  Changing 
integration variables to $x=q^{2}$ we find
\be
I=\tilde{I}_{0}' h^{d/2}\int_{0}^{\infty} dx x^{d/2-1}
e^{2(x-x^{2})t-\frac{h}{2}x}
\ee 
where $\tilde{I}_{0}' =\tilde{I}_{0}/2$.
The leading large time dependence can be extracted from
the integral by completing the square in the argument
of the exponential or using the stationary phase method.
We find, to leading order in large t:
\be
I=\tilde{I}_{0}\left(\frac{b}{2}\right)^{d/2-1}
\sqrt{\frac{\pi}{2t}}e^{\phi^{2}}
\label{eq:195}
\ee
where
$b=1-\frac{h}{4t}$, $\phi^{2} = \frac{t}{2}b^{2}$.
From this result for $I$ we see that
$\dot{I}=\omega I$
where
\be
\omega=(\frac{d}{2}-1)\frac{\dot{b}}{b}
+2\phi\dot{\phi}-\frac{1}{2t}
=\frac{1}{2}-\frac{1}{2t}+
{\cal O}(t^{-2})
~~~.
\label{eq:199}
\ee
Going further we have
$\ddot{I}=\omega^{2} I +\dot{\omega}I$
which leads easily to the useful result:
\be
\frac{\ddot{I}}{I}-\left(\frac{\dot{I}}{I}\right)^{2}
=\dot{\omega}
~~~.
\ee
Turning to the evaluation of $J$ given by Eq.(\ref{eq:160}),
using Eq.(\ref{eq:195}), we find
\be
J=\frac{1}{2dt}\left[\frac{(d/2 -1)}{b}
+2\phi\frac{\partial \phi}{\partial b}\right]I
~~~.
\ee
Working to leading order in time we find
$J=\frac{I}{2d}$.
Putting all of this together in Eq.(\ref{eq:163})
\be
\bar{v}^{2}=\frac{\dot{\omega}}{J/I}=\frac{d}{t^{2}}
\ee
since, from Eq.(\ref{eq:199}),
$\dot{\omega} =\frac{1}{2t^{2}} +{\cal O}(t^{-3})$.
The final result for $\bar{v}^{2}$ is independent of the
initial conditions.  We see that the COP average vortex
speed is qualitatively slower than the NCOP case:
\be
\frac{\bar{v}^{2}_{COP}}{\bar{v}^{2}_{NCOP}}\approx \frac{1}{t}
~~~.
\ee
The computation of $\bar{v}_{\mu}$ using Eq.(\ref{eq:163})
has been checked numerically in the simplest
$n=d=2$ case where $I$ can be evaluated explicitly in
terms of an erfc function.

\section{Conclusions}

We have presented here the detailed calculation
of the vvpdf including the time dependent vortex
scaling velocity $\bar{v}_{\mu}$  for a class of
models beyond the original nonconserved
TDGL models. The class of models studied includes
the conserved TDGL model and certain anisotropic
models.  In the conserved case it is found that
the average vortex speed falls off as $t^{-1}$
compared to the NCOP case where $\bar{v}\approx t^{-1/2}$. 
It is our intension to numerically test the predictions
for the COP case.

We see that there is
self-consistent confirmation that in dealing with
vortex velocities one can organize things in terms of
averages over an auxiliary Gaussian field.  We require self-
consistently that this field and the order parameter
field share the same zeros.  A similar development
can be worked out for string defects\cite{MW98,MAZ99}.

Acknowledgements: This work was supported
by the National Science Foundation under Contract
No. DMR-0099324.

\pagebreak

\end{document}